\pdfoutput=1
\documentclass[runningheads]{llncs}

\usepackage{hyperref}
\usepackage{mathtools}
\usepackage{amsmath}
\usepackage{amstext}
\usepackage{tikz}
\usepackage{listings}
\usepackage{tikz-3dplot}
\usepackage{graphicx}
\usepackage{caption}
\usepackage{subcaption}
\usepackage{pgfplots}
\usepackage{pgfplotstable}
\usepackage{array}
\usepackage{xcolor}
\usetikzlibrary{patterns}
\pgfplotsset{compat=1.7}

\usepackage{wrapfig} 

\tdplotsetmaincoords{70}{120} 
\tdplotsetrotatedcoords{0}{0}{90} 

\begin{document}

\def\xlist{4}
\def\ylist{4}

\newcommand{\editrevA}[1]{{\color{black}#1}}
\newcommand{\editrevB}[1]{{\color{black}#1}}
\newcommand{\editrevC}[1]{{\color{black}#1}}
\newcommand{\editrevD}[1]{{\color{black}#1}}

\newcommand{\fillrandomly}[6]{
  \pgfmathsetmacro\diameter{#5*2}
  \draw [very thin,black] (#1,#2) rectangle (#3,#4);
  \foreach \i in {1,...,#6}{
      \pgfmathsetmacro\x{#1+#5+rnd*(#3-#1-2*#5)}
      \pgfmathsetmacro\y{#2+#5+rnd*(#4-#2-2*#5)}
      \xdef\collision{0}
      \foreach \element [count=\i] in \xlist{
        \pgfmathtruncatemacro\j{\i-1}
        \pgfmathsetmacro\checkdistance{ sqrt( ({\xlist}[\j]-(\x))^2 + ({\ylist}[\j]-(\y))^2 ) }
        \ifdim\checkdistance pt<\diameter pt
          \xdef\collision{1}
          \breakforeach
        \fi
      }
      \ifnum\collision=0
        \xdef\xlist{\xlist,\x}
        \xdef\ylist{\ylist,\y}
        \fill [red, thick] (\x,\y) circle [radius=#5];
      \fi

    }
}

\title{Kokkos-Based Implementation of MPCD on Heterogeneous Nodes}

\author{Rene Halver\inst{1}\orcidID{0000-0002-4895-3762} \and
  Christoph Junghans\inst{2}\orcidID{0000-0003-0925-1458} \and
  Godehard Sutmann\inst{1,3}\orcidID{0000-0002-9004-604X}}

\institute{J\"ulich Supercomputing Centre, Institute for Advanced Simulation,
  Forschungszentrum J\"ulich, 52425 J\"ulich, Germany \email{r.halver@fz-juelich.de, g.sutmann@fz-juelich.de} \and
  Los Alamos National Laboratory, CCS-7, 87545 Los Alamos, New Mexico, USA \email{junghans@lanl.gov} \and
  ICAMS, Ruhr-University Bochum, 44801 Bochum, Germany}



\maketitle

\begin{abstract}
  The Kokkos based library Cabana, which has been developed in the Co-design Center for Particle Applications (CoPA), is used for the implementation of Multi-Particle Collision Dynamics (MPCD),
  a particle-based description of hydrodynamic interactions. It allows a performance portable implementation, which has been used to study the interplay between CPU and GPU usage on a multi-node system. As a result, we see most advantages in a homogeneous GPU usage, but we also discuss the extent to heterogeneous applications, using both CPU and GPU concurrently.\\
  %
\end{abstract}

\section{Introduction}

The recent development of high-end parallel architectures shows a clear trend to a heterogeneity of compute components, pointing towards a dominance of General Purpose Graphics Processing Units (GPU) as accelerator components, compared to the Central Processing Units (CPU). According to the Top~500 list~\cite{Dongarra2011}, more than 25\% of the machines have GPU support while the overall performance share is more than 40\%, i.e., heterogeneous cluster architectures have a large impact for high compute performance.  Often \editrevA{these} nodes consist of only a few multicore CPUs, while supporting 2-6 GPUs. In many applications one can observe a trend that the most powerful component of the nodes, i.e. the GPUs, is addressed, while the CPUs are used as administrating or data management components. A reason might be the additional overhead
in writing/maintaining two different code versions for each architecture, as usually a CPU code cannot simply run on a GPU or vice versa.


With the advent of performance portable programming models, such as Kokkos~\cite{CarterEdwards20143202}
or Raja~\cite{RAJAwebsite} it has become possible to use the same code base for different
architectures, most prominently including CPUs or GPUs. It might be tempting to use the full capacity of a compute-node concurrently, i.e. not wasting compute resources because of the disparate character of the architecture and programming model. In this case one encounters both different performance characteristics of components and possibly a non-negligible data transfer between components. This discrepancy might be targeted by load balancing strategies which would need to take into account hardware and software specific characteristics to achieve an overall performance gain.

In the present paper we consider a stochastic particle based method for the simulation of hydrodynamic phenomena,  i.e. the Multi-Particle Collision Dynamics (MPCD)~\cite{Gompper}
algorithm and its implementation with Cabana~\cite{Cabana,Mniszewski_2021,slattery.s.2022a},
a Kokkos based library. We first introduce the underlying MPCD method and then describe the Cabana library. We then present some benchmark results and finally draw conclusions from our findings and give some outlook for further research.


\section{Multi-Particle Collision Dynamics}
\label{sec:MPCD}

MPCD is a particle-based description for hydrodynamic interactions in an incompressible
fluid. The method is based on a stochastic collision scheme in which particles, that describe the
simulated fluid are rotated in velocity space while conserving linear momentum and energy (variants exist which also conserve angular momentum~\cite{Gompper}). The method proceeds by sorting particles into a regular mesh with grid cells of size of a characteristic length scale. In order to transport momentum and energy across the system, the mesh is randomly shifted in each time step, changing the local environment of each particle stochastically.
For each particle in a cell its relative velocity with respect to the center-of-mass ({\em com}) velocity of the cell is computed. This velocity is split into a parallel and perpendicular component with respect to a randomly oriented axis in the cell.  Consequently, the perpendicular component is rotated around that axis by a fixed angle, which determines together with the particle mass and density, the time step and the cell length the diffusion and viscosity of the fluid under consideration. This procedure can be shown to mimic hydrodynamic behaviour and, in a limiting case, enters into the Navier Stokes equations~\cite{Gompper}. Using this procedure the conservation of linear momentum and energy is guaranteed and can also be coupled to embedded particles, simulated by other methods, e.g. molecular dynamics, thereby coupling particle dynamics to a hydrodynamic medium~\cite{Gompper,huang.c.c.2010b}.

From an algorithmic point of view, three main parts can be identified, i.e. (i) the local identification of particles in the underlying cell structure and the computation of {\em com} velocities of cells; (ii) the computation of the relative velocities of particles with respect to the {\em com} velocity of a cell; (iii) rotation of perpendicular velocity component of particles around a random axis. These parts will be discussed separately in Sec.~\ref{sec:cabana} in more detail in the context of the Cabana implementation.



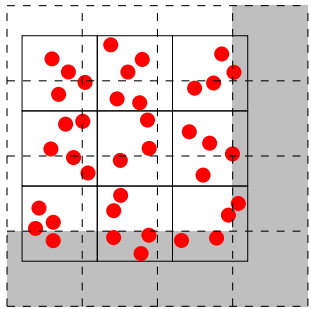
\begin{figure}
  \centering
  \begin{tikzpicture}
    \pgfmathsetseed{2}

    \foreach \x in {-0.2,...,2.8}
      {
        \fill [color=lightgray] (\x,-0.6) rectangle (\x+1,0.4);
      }

    \foreach \y in {-0.6,...,2.4}
      {
        \fill [color=lightgray] (2.8,\y) rectangle (3.8,\y+1);
      }

    \foreach \x in {0,...,2} {
        \foreach \y in {0,...,2} {
            \fillrandomly{\x}{\y}{\x+1}{\y+1}{0.1}{6}
          }
      }

    \foreach \i in {-0.2,...,3.8} {
        \draw [thin,black,dashed] (\i,-0.6) -- (\i,3.4);
      }
    \foreach \i in {-0.6,...,3.4} {
        \draw [thin,black,dashed] (-0.2,\i) -- (3.8,\i);
      }

  \end{tikzpicture}
  \caption{ \label{fig:MPCD_grid} Illustration of the shifted collision cell grid (black, dashed) in comparison to the static
    logical cell grid (black, solid). The grey cells mark the periodic images of the shifted grid.}
\end{figure}

\section{Implementation with Cabana}
\label{sec:cabana}

The aim of the implementation was to write a code, that is performance portable between clusters consisting of
CPU and clusters with GPU nodes, which often consist of one or two CPUs and a number of GPUs ranging from
two to six. \editrevA{Maintaining} two or more codebases for all targeted architectures increases the overhead time of, e.g., design or maintenance time,
and calls for solutions which allow a unified approach for various architectures.

For this reason performance portable programming models are attractive for reducing time spent with porting codes to various architectures. One of the more popular programming models in \editrevA{this} regard is Kokkos~\cite{CarterEdwards20143202},
which provides an abstraction layer for data structures, called \textit{Views}, while providing different \textit{ExecutionSpaces}
which can either be on the host (usually the CPU) or on devices, i.e. GPUs or other accelerator cards, e.g. Intel KNLs.
Kokkos uses different backends to provide this performance portability, e.g. CUDA for the use of NVIDIA GPUs or ROCm for the
use of AMD GPUs. Furthermore, OpenMP or PThread backends can be used among others to utilise multicore architectures of CPUs.\\
Within the Exascale Computing Project (ECP)~\cite{ECP} funded by the Department of Energy (DoE) in the \editrevA{USA}, the Co-Design Center
for Particle Applications (CoPA)~\cite{CoPA} developed a performance portable library, based on Kokkos, with the main focus of supporting
the development of particle and grid based codes on HPC systems. Cabana not only provides data structures based on
Kokkos \textit{Views} but also provides routines in order to facilitate data transfer between different processes in a distributed-memory
environment, based on MPI.\\
Since the MPCD method is a mixture of a particle and a grid based method (due to the requirement to sort the particles into cells),
the implementation of the MPCD code using Cabana was considered reasonable. In the rest of the section the main points of the implementation
will be presented.

\subsection{Collection of Particles in Cells}
\label{ssec:linked_cell}

Before the {\em com} velocity for a cell can be calculated, it is necessary to identify the particles that reside in each
collision cell. One technique to achieve this is the linked-cell list. Accordingly, all particles are checked and flagged with a
cell identifier to which they belong to. In addition, a (linked) list of particles belonging to the cell is created. Listing \ref{lst:linked_cell_creation}
shows how such a list is created in Cabana. The use of Cabana simplifies the creation of such a linked cell list, as Cabana deals with the
issues of creating a linked cell list in a multithreaded environment, as described e.g. in \cite{Halver:279249} or \cite{8345421}.

\begin{lstlisting}[basicstyle=\footnotesize\ttfamily,language=C++,caption={Creation of the linked cell list of the shifted collision cell grid},label=lst:linked_cell_creation]
  // boundaries of spacial domains
  double gridMin[3], gridMax[3];
  for (int d = 0; d < 3; ++d)
  {
    gridMin[d] = domBorders(2*d) - (double)haloWidth 
                 * cellSize(d) + offset(d);
    gridMax[d] = domBorders(2*d+1) + (double)haloWidth 
                 * cellSize(d) + offset(d);
  }
  // creating the linked cell list
  // r = list of particle positions
  // cellSize = size of linked cells (3d)
  Cabana::LinkedCellList<DeviceType> 
    linkedList( r, cellSize, gridMin, gridMax );
  // permute the particle AoSoA to correspond to the cells
  Cabana::permute( linkedList, particles );
\end{lstlisting}

\subsection{Communication of Required Information}
\label{ssec:communication}

As described in section~\ref{sec:MPCD}, it is necessary to compute the {\em com} velocity, i.e. the velocity in a zero momentum frame with regard to the local collision cell \cite{goldstein.h.2002a}, in order to
calculate the collisions within each mesh cell, which requires all velocities and masses of particles that reside within the given collision
cell. The underlying parallel algorithm is based on a domain decomposition, where compute resources administrate geometrical spatial regions which are connected. Since the underlying mesh is shifted in each time step cells might be split among several domains. To compute a unique value for the {\em com} velocity, one can either collect all particles together with their properties on a local domain or one can compute the partial {\em com} velocities on each local domain and then reduce this value among those processes which share the given cell.

The first of these methods has the advantage that since all particles are collected on a single domain, the computation of the
  {\em com} velocity and the following rotation of velocities can be executed without the need of additional communication steps in between.
The disadvantage is that it requires the communication of particle data in each time step, since the collision cell mesh
needs to be shifted in each time step to avoid artefacts in the computation of the hydrodynamic interactions. Listing \ref{lst:comm_cabana_particles}
shows the necessary steps to prepare the particle migration between domains. \editrevD{Shown here is a way to try to avoid unnecessary branching while determining the
  target processes for particles. This is done by masking the target processes with a base-3 number, where each 'bit' indicates either a shift down($0$) or up ($1$) or
  residing in the domain's boundary concerning that Cartesian direction. As an example a base-3 number of $(201)_3$ would be assigned to a particle leaving the local domain
  in positive x-direction and negative y-direction, while stay in the same z-region, as the local domain. This way to determine target processes should improve execution
  on GPU, with the tertiary operator being removed, in case that true is cast to integer one and false to integer zero.}

\begin{lstlisting}[basicstyle=\footnotesize\ttfamily,language=C++,caption={Particle based communication with Cabana},label=lst:comm_cabana_particles]
Kokkos::parallel_for(
  Kokkos::RangePolicy<ExecutionSpace>(0, nParticles), 
  KOKKOS_LAMBDA (const size_t i)
  {
    int dims = 1, index = 0;
    // compute the direction of the neighbour the particle 
    // needs to be moved to and use dims to compute a 
    // base 3 mask:
    // (xyz)_3 with 0 (left), 1 (remains), 2 (right)
    // r = list of particle positions
    for (int d = 2; d >= 0; --d)
    {
        index += dims * 
          ( 1 - ((r(i,d) < domBorders(2*d))?1:0) + 
          ((r(i,d) >= domBorders(2*d+1))?1:0) );
        dims *= 3;
    }
    // tag the particle with the target neighbour rank
    export_ranks(i) = neigs(index);
  });
Kokkos::fence();

// create particles distribution object and 
// migrate particles to targets
Cabana::Distributor<DeviceType> dist( mpiCart, 
                      export_ranks, neighbours );
Cabana::migrate(dist, particles);
\end{lstlisting}

In contrast, the second method allows the use of a stable, halo-based communication scheme, where particles are not
necessarily communicated in each time step, but only when leaving a halo region around the local domain, allowing the distributed computation
of partial {\em com} velocities, that are reduced with a static communication scheme. The result is then sent back to the domains
sharing the same cell. Listing \ref{lst:comm_cabana_grid} shows the required function calls to Cabana to
do the halo exchange. This work, related to mesh administration, is implemented in {\em Cajita}, which  is part of Cabana. In addition, it provides methods for particle-grid interactions,
e.g. interpolation of particle properties to a grid, which is, however, not used in this work. Furthermore, Cajita provides a domain-based load balancing based on a tensor decomposition scheme, provided by the ALL library~\cite{ALL-web}.

\begin{lstlisting}[basicstyle=\footnotesize\ttfamily,language=C++,caption={Grid based halo communication with Cabana},label=lst:comm_cabana_grid]
  // create the halo communication object based 
  // on the Cajita grid 
  auto arrHalo = Cajita::createHalo( *arrNode, 
                 Cajita::NodeHaloPattern<3>());
  // [...] computation of com velocities
  // bring the data to the halo cells
  arrHalo->gather(ExecutionSpace(), *arrNode);
  // collect the data from the halo cells
  arrHalo->scatter(ExecutionSpace(), 
       Cajita::ScatterReduce::Sum(), *arrNode);
\end{lstlisting}

For the implementation of the two different communication schemes two different kinds of communication in Cabana were used. For the former
method, the particle-based one, Cabana provides a \textit{Distributor} class, which allows the transfer of particle data between processes.
This requires that particles are tagged with the target process, so that the \textit{Distributor} object can generate a communication
topology for this specific transfer. As a consequence this object needs to be recreated in every time step, since the communication
pattern in each time step changes due the random shift of the collision cell grid and particle movements across domain borders. \\

For the second communication pattern, reducing the partial results and redistributing them, a halo-based communication on a grid is
used. For this purpose, two different grids are combined, i.e. a logical collision grid which is used for communication and a linked-cell list, which
sorts the particles into the shifted collision cell grid. Since the number and size of mesh cells in each grid is identical, both grids can be perfectly matched onto
each other. The particles are sorted into the linked-cell list (Sec.~\ref{ssec:linked_cell}) from where the {\em com} momentum of each cell is computed. For collision cells, overlapping with domain \editrevA{borders} (Fig.~\ref{fig:MPCD_grid}),
a halo-based communication reduces the partial results on the process which administrates the logical cell. This process \editrevA{redistributes}
\editrevA{the} reduced sum back to each participating neighbour, where the rotations of velocities are computed for residing particles. Since the number of cells is usually
far smaller than the number of particles, this leads to (i) a static communication scheme (for each iteration step the same operations on the same amount of data) and (ii) a reduced and constant amount of data that needs to
be communicated.\\

During the development, it became apparent that the second communication scheme leads to a better performance due to the reduced amount of transferred data
and the strongly reduced necessity to recreate communication patterns, due to the stable communication scheme of the halo exchange (this needs to be done only once in the beginning or after possible load balancing steps, after which the communication pattern is static). In
addition, the transfer of particles can be reduced to cases, where particles left the halo region surrounding the local domain, instead of
being required in every time step.

\subsection{Rotation of Velocities}

To simplify the computation of the velocity rotation, the linked cell list mentioned in section \ref{ssec:linked_cell}
is used to sort particles into the correct cell of the collision cell grid. Using the {\em com} velocity, gathered by one of the two
previously described methods, the linked-cell list provides the particles which belong to the given cell and their velocity vector rotated.

\begin{lstlisting}[basicstyle=\footnotesize\ttfamily,language=C++,caption={Using the linked cell list from listing \ref{lst:linked_cell_creation} to compute the com velocity},label=lst:linked_cell_use]
  // Kokkos parallel_for iterates over  
  // all cells on local domain
  // vcm = Kokkos::View containg the center
  //       of mass velocites for each
  //       collision cell 
  // v   = Cabana::slice containing
  //       particle velocities
  // m   = Cabana::slice containing
  //       particles masses  
  Kokkos::parallel_for(Kokkos::RangePolicy<ExecutionSpace>
    (0, linkedList.totalBins()), 
    KOKKOS_LAMBDA( const size_t i)
    {
      int ix, iy, iz;
      // computing the cartesian coordinates of the cell
      linkedList.ijkBinIndex(i, ix, iy, iz);
      int binOff = linkedList.binOffset(ix, iy, iz);
      // compute com velocity
      for (int d = 0; d < 4; ++d)
        vcm(ix,iy,iz,d) = 0.0;
      // computing com momentum and sum of mass
      for (int n = 0; n < linkedList.binSize(ix,iy,iz); ++n)
      {
        for (int d = 0; d < 3; ++d)
          vcm(ix,iy,iz,d) += v(binOff + n, d) * 
                             m(binOff + n);
        vcm(ix,iy,iz,3) += m(binOff + n);
      }
    });
  Kokkos::fence();
\end{lstlisting}

\section{Benchmarks and Discussion}
\label{sec:benchmarks}

\begin{figure*}[ht]
  \centering
  \resizebox{\columnwidth}{!}
  {
    \begin{tabular}{c c c}

      \begin{subfigure}[pt]{0.45\linewidth}
        \begin{tikzpicture}
          \begin{axis} [xlabel=system size,
              ylabel=runtime,
              xmode=log,
              ymode=log,
              ymin=1,
              ymax=20000,
              width=5.9cm, height=5.9cm,
              legend style = {at={(0.43,0.15)},anchor=west, nodes={scale=0.5, transform shape}},
              xtick={32,64,128,256,512},
              xticklabels={32,64,128,256,512}]
            \addplot[color=blue, mark=o] coordinates{
                (32, 5.3651)
                (64, 49.49)
                (128, 819.06)
              };
            \addplot[color=red, mark=x] coordinates{
                (32, 2.75163)
                (64, 11.2)
                (128, 132.057)
                (256, 1935.45)
                (512, 19354.5)
              };
            \addplot[color=black, dashed] coordinates{
                (32, 5.3651)
                (64, 8*5.3651)
                (128, 64*5.3651)
                (256, 512*5.3651)
                (512, 4096*5.3651)
              };
            \addplot[color=black, dotted, thick] coordinates{
                (32, 2.75163)
                (64, 8*2.75163)
                (128, 64*2.75163)
                (256, 512*2.75163)
                (512, 4096*2.75163)
              };
            \legend{1 node,4 nodes, \editrevA{ideal} (1 node), \editrevA{ideal} (4 nodes)}
          \end{axis}
        \end{tikzpicture}
        \caption{CPU (Fortran)}
        \label{fig:CPUperformance}
      \end{subfigure}

       &

      \begin{subfigure}[pt]{0.45\linewidth}
        \begin{tikzpicture}
          \begin{axis} [
              xlabel=system size,
              xmode=log,
              ymode=log,
              ymin=1,
              ymax=20000,
              width=5.9cm, height=5.9cm,
              legend style = {at={(0.45,0.2)},anchor=west, nodes={scale=0.5, transform shape}},
              xtick={32,64,128,256,512},
              xticklabels={32,64,128,256,512}]
            \addplot[color=blue, mark=o] coordinates{
                (32, 52.9649)
                (64, 75.5525)
                (128, 162.88)
                (256, 652.249)
                (512, 5897.44)
              };
            \addplot[color=brown, mark=square*] coordinates{
                (32, 59.4714)
                (64, 86.4785)
                (128, 130.268)
                (256, 335.118)
                (512, 2774.35)
              };
            \addplot[color=red, mark=x] coordinates{
                (32, 53.3625)
                (64, 78.7678)
                (128, 107.288)
                (256, 211.506)
                (512, 1429.14)
              };
            \addplot[color=black, mark=otimes*] coordinates{
                (128, 92.144)
                (256, 182.184)
                (512, 648.252)
              };
            \addplot[color=gray, mark=triangle*] coordinates{
                (256, 144.602)
                (512, 327.49)
              };
            \addplot[color=black, dashed] coordinates{
                (32, 52.9649)
                (64, 8*52.9649)
                (128, 64*52.9649)
                (256, 512*52.9649)
                (512, 4096*52.9649)
              };
            \legend{1 node,2 nodes,4 nodes,8 nodes,16 nodes, \editrevA{ideal} (1 node)}
          \end{axis}
        \end{tikzpicture}
        \caption{GPU (C++)}
        \label{fig:GPUperformance}
      \end{subfigure}

       &

      \begin{subfigure}[pt]{0.45\linewidth}
        \begin{tikzpicture}
          \begin{axis} [xlabel=system size,
              xmode=log,
              ymode=log,
              ymin=1,
              ymax=3000,
              width=5.9cm, height=5.9cm,
              legend style = {at={(0.54,0.15)},anchor=west, nodes={scale=0.5, transform shape}},
              xtick={32,64,128,256},
              xticklabels={32,64,128,256}]
            \addplot[color=blue, mark=o] coordinates{
                (32, 5.3651)
                (64, 49.49)
                (128, 819.06)
              };
            \addplot[color=red, mark=x] coordinates{
                (32, 2.75163)
                (64, 11.2)
                (128, 132.057)

              };
            \addplot[color=black, mark=otimes*] coordinates{
                (32, 42.6227)
                (64, 266.971)
                (128, 2268.8)

              };
            \addplot[color=gray, mark=square*] coordinates{
                (32, 16.3708)
                (64, 78.7868)
                (128,555.418)
              };
            \legend{F90 1N, F90 4N, C++ 1N, C++ 4N}
          \end{axis}
        \end{tikzpicture}
        \caption{F90 (s) and C++ (OMP) }
        \label{fig:CPUCPUperformance_comparison}
      \end{subfigure}

      \\[3cm]

      \begin{subfigure}[pt]{0.45\linewidth}
        \begin{tikzpicture}
          \begin{axis} [xlabel=system size,
              ylabel=runtime,
              xmode=log,
              ymode=log,
              ymin=1,
              ymax=3000,
              width=5.9cm, height=5.9cm,
              legend style = {at={(0.54,0.15)},anchor=west, nodes={scale=0.5, transform shape}},
              xtick={32,64,128,256},
              xticklabels={32,64,128,256}]
            \addplot[color=blue, mark=o] coordinates{
                (32, 5.3651)
                (64, 49.49)
                (128, 819.06)
              };
            \addplot[color=red, mark=x] coordinates{
                (32, 2.75163)
                (64, 11.2)
                (128, 132.057)

              };
            \addplot[color=black, mark=otimes*] coordinates{
                (32, 38.3348)
                (64, 221.384)
                (128, 1967.42)

              };
            \addplot[color=gray, mark=square*] coordinates{
                (32, 15.0551)
                (64, 69.0582)
                (128, 474.669)
              };
            \legend{F90 1N, F90 4N, C++ 1N, C++ 4N}
          \end{axis}
        \end{tikzpicture}
        \caption{F90 (s) and C++ (serial) }
        \label{fig:CPUCPUperformance_comparison2}
      \end{subfigure}

       &

      \begin{subfigure}[pt]{0.45\linewidth}
        \begin{tikzpicture}
          \begin{axis} [xlabel=system size,
              xmode=log,
              ymode=log,
              width=5.9cm, height=5.9cm,
              legend style = {at={(0.0,0.9)},anchor=west, nodes={scale=0.5, transform shape}},
              xtick={32,64,128,256},
              xticklabels={32,64,128,256}]
            \addplot[color=blue, mark=o] coordinates{
                (32, 5.3651)
                (64, 49.49)
                (128, 819.06)
              };
            \addplot[color=red, mark=x] coordinates{
                (32, 65.6)
                (64, 86.166)
                (128, 174.175)
              };
            \legend{CPU,GPU}
          \end{axis}
        \end{tikzpicture}
        \caption{F90 (CPU) C++ (GPU)}
        \label{fig:GPUCPUperformance_comparison}
      \end{subfigure}

       &

      \begin{subfigure}[pt]{0.45\linewidth}
        \begin{tikzpicture}
          \begin{axis} [xlabel=system size,
              xmode=log,
              ymode=log,
              ymin=10,
              ymax=2000,
              width=5.9cm, height=5.9cm,
              legend style = {at={(0.0,0.9)},anchor=west, nodes={scale=0.5, transform shape}},
              xtick={32,64},
              xticklabels={32,64}]
            \addplot[color=blue, mark=o] coordinates{
                (32, 58.9619)
                (64, 237.485)
              };
            \addplot[color=red, mark=x] coordinates{
                (32, 52.9649)
                (64, 75.5525)
              };
            \legend{GPU+CPU,GPU}
          \end{axis}
        \end{tikzpicture}
        \caption{CPU+GPU/GPU (C++)}
        \label{fig:GPUCPUperformance}
      \end{subfigure}
    \end{tabular}
  }
  \caption{Performance comparison between existing Fortran implementation and new Cabana implementation using multiple nodes.}
  \label{fig:benchmarks}
\end{figure*}
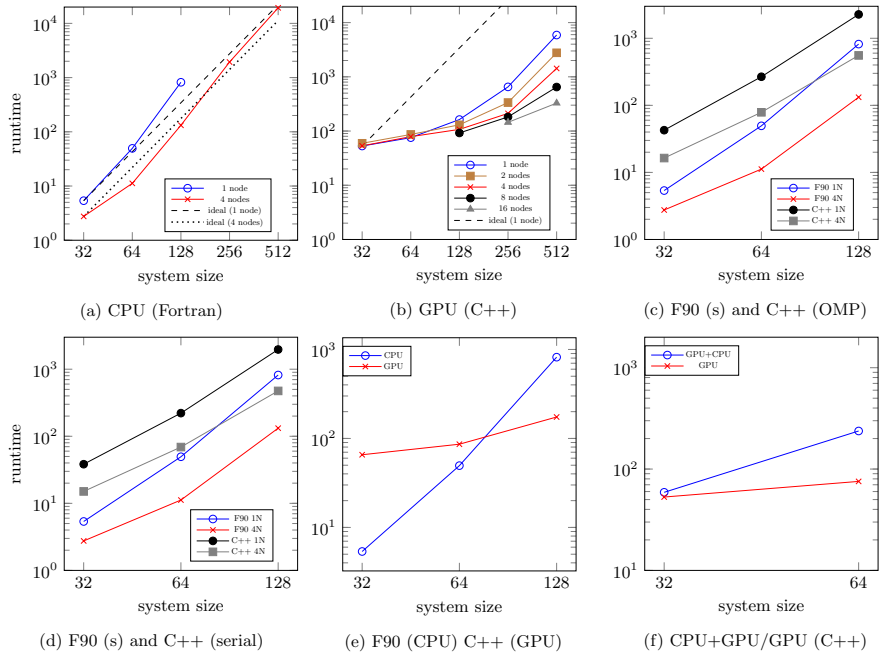

\begin{table}[htbp]
  \centering
  \caption{Tables of runtimes for the different implementations. \editrevA{Empty cells indicate combinations of node numbers and system sizes, for
      which additional measurements did not show additional information. All runtimes presented are given in seconds.}}
  \resizebox{\columnwidth}{!}
  {
    \begin{tabular}{c}
      \begin{subtable}[h]{.95\textwidth}
        \centering
        \small
        \caption{Runtimes for GPU C++ variant, using 4 GPUs on each node.}
        \begin{tabular}{|c|c|c|c|c|c|}
          \hline
          size & 1 node  & 2 nodes & 4 nodes & 8 nodes & 16 nodes \\
          \hline
          32   & 52.96   & 59.47   & 53.36   &         &          \\
          64   & 75.55   & 86.48   & 78.77   &         &          \\
          128  & 162.88  & 130.27  & 107.29  & 92.11   &          \\
          256  & 652.25  & 335.12  & 211.51  & 182.18  & 144.60   \\
          512  & 5897.44 & 2774.35 & 1429.14 & 648.252 & 327.49   \\
          \hline
        \end{tabular}
        \label{tab:GPUruntimes}
      \end{subtable}
      \\
      \begin{subtable}[h]{.95\textwidth}
        \centering
        \small
        \caption{Runtimes for CPU variants, Fortran (F90) and the C++ based variants, i.e.
          OpenMP-based (OMP) or serial, i.e. no hybrid parallelization, using one or four nodes (N). OMP uses 8 MPI ranks with 6 threads each on a node, the Fortran and serial version 48 MPI ranks per node.
          \editrevB{Only system sizes up to edge length 128 are presented due to the longer runtimes.}}
        \label{tab:CPUruntimes}
        \begin{tabular}{|c|c|c|c|c|c|c|}
          \hline
          size & F90 1 N & F90 4 N & OMP 1 N & OMP 4 N & serial 1 N & serial 4 N \\
          \hline
          32   & 5.36    & 2.75    & 42.62   & 16.37   & 38.33      & 15.06      \\
          64   & 49.59   & 11.20   & 284.73  & 78.79   & 221.38     & 69.06      \\
          128  & 819.06  & 132.06  & 2268.80 & 555.42  & 1967.42    & 474.67     \\
          \hline
        \end{tabular}
      \end{subtable}
    \end{tabular}
  }
  \label{tab:runtimes}
\end{table}
\raggedbottom
For the benchmark runs simple fluid systems were used, i.e. a pure MPCD fluid in 3d periodic boundary conditions.
Each cubic collision cell has an edge length of one length unit, while containing $\langle N_c \rangle=10$ particles on average. Each system in the benchmarks is cubic with side length $L$ (the edge length $L$ given as the system size in the following graphs, i.e. Fig.~\ref{fig:benchmarks}), from where the total number of particles in a system is computed as $N=L^3\, \langle N_c \rangle$.
To check the performance of the newly implemented code, it was compared to an existing Fortran implementation of the MPCD algorithm~\cite{huang.c.c.2010b,MP2Cwebsite}.\\
%
The benchmarks were performed on the Juwels booster module~\cite{JUWELS} at J\"ulich Supercomputing Centre, consisting of GPU nodes with four NVIDIA A100 cards and two AMD EPYC 7402 processors, with 24 cores each. To maintain comparability of the benchmarks
the pure CPU runs were also performed on these nodes. \editrevB{Since the GPU nodes are much more powerful in their computing capabilities, we performed the benchmarks for the GPU
  runs on node numbers from one to 16, doubling the node count each time. For the CPU, expecting longer runtimes we chose to compare single node runs with runs on four nodes, while also restricting the
  system size to a maximum edge length of 128 while for the GPU runs we performed the benchmarks to a maximum edge length of 512. The edge length directly influences the number of particles in the
  simulation, since there are about $l^3$ collision cells in the system, with $l$ being the edge length of the system, each collision cell containing ten particles on average.}\\
\editrevC{As backends for Kokkos were the AMD and Ampere70 used, since these corresponded best to the available hardware. No further optimization on the basis of compiler flags was
  attempted yet due to time constraints, but these tests will be performed in the future.}
Table~\ref{tab:runtimes} and Fig.~\ref{fig:benchmarks} show results for four different benchmarks: (i) C++/Kokkos implementation with GPU variant (Table~\ref{tab:GPUruntimes} and
Fig.~\ref{fig:GPUperformance}); (ii) C++/Kokkos variant with OpenMP (Table~\ref{tab:CPUruntimes}
and Fig.~\ref{fig:CPUCPUperformance_comparison}); (iii) C++/Kokkos variant with
serial backend and (iv) the previous implementation of the MPCD algorithm in Fortran
(Table~\ref{tab:CPUruntimes} and Fig.~\ref{fig:CPUperformance}) for comparison with the new implementation. \\
The original Fortran code shows a quite good scaling behaviour for all studied cases (edge lengths $L\in [32,512]$), as
can be seen in Fig.~\ref{fig:CPUperformance}. In comparison to that the scaling behaviour
of the GPU variant of the C++ implementation shows for the smaller system sizes a
super-linear scaling behaviour, before reaching linear behaviour at system sizes
$256$ and $512$, indicating that smaller sizes not fully utilise the GPU (Fig.~\ref{fig:GPUperformance}). \\
When comparing the performance of the Fortran implementation (Fig.~\ref{fig:CPUperformance}) and the CPU based variants
of the C++ version, i.e. OpenMP based or serial, it can be seen that Fortran
achieves much better results (Figs.~\ref{fig:CPUCPUperformance_comparison},\ref{fig:CPUCPUperformance_comparison2}). An explanation for
this behaviour still needs to be analysed in more depth. But first results point towards a different level of optimization (which is not the main focus of this article). In contrast, the GPU
variant is able to outperform the Fortran implementation given sufficiently large system
sizes, as can be seen in Fig.~\ref{fig:GPUCPUperformance_comparison}, comparing the benchmark
results on a single node, respectively. \editrevC{Here only the results for system sizes 32 and 64 are shown, since the measurement strongly hint that for larger system sizes the gap between
  hybrid execution and pure GPU execution will only widen.} \\ \indent
Furthermore, it was tested on a single node if the combination of GPU and CPU could result
in a better performance than only GPU computations. Due to the obtained performance of the
CPU-based C++ variants, the results indicate at this stage
no performance gain for hybrid execution (Fig.~\ref{fig:GPUCPUperformance}). In case of a performance improvement of the CPU-based variants, this result might change for smaller system sizes.
Note that for small systems load balancing GPU and CPU ranks can improve the overall performance for hybrid execution significantly, but not sufficiently in order to outperform either pure CPU or GPU. This does not lead to a recommendation of a hybrid execution model at this stage.

\section{Conclusion and Outlook}

Considering the benchmark results of the new implementation of the MPCD code the following conclusions can be drawn:

(i) \editrevD{It} is possible to implement a scalable MPCD algorithm with Cabana, that for large enough systems
is faster on GPUs than the existing Fortran implementation.
The CPU variant of the Cabana implementation
needs to be improved upon to bring the performance closer to the one of the Fortran code.

(ii) Load balancing between CPU and GPU can support hybrid execution, but was not found to increase performance beyond the one of pure CPU or GPU usage.

(iii) The porting effort from a pure CPU variant to a multi-architecture variant was significantly decreased by using Cabana, which offers an architecture independent development and code implementation which provides a unified and transparent view for the programmer. Porting effort is therefore dramatically reduced by maintaining performance (which was not the focus here, but which is demonstrated for other use cases ~\cite{reuter.k.2020a,CarterEdwards20143202,Halver:885401}).

(iv) The implementation of the MPCD algorithm allows further investigation of coupled simulations of MPCD
fluids with embedded Molecular Dynamics (MD) systems, e.g. polymer chains. For this, an implementation based on a unified formulation of MD and MPCD, as described, e.g., in ~\cite{Gompper,huang.c.c.2010b}, is required. Since the ratio of MD- to MPCD particles is often small, this could profit from a hybrid implementation and execution model, which invites to further investigations, including execution models for modular supercomputing.


\bibliographystyle{splncs04}
\bibliography{CP60}

\end{document}